\documentclass[twoside,epsf]{article}
\usepackage{graphicx}
\newcommand{\lascia}[1]{}

\newcount\Mac  \Mac=0  % devo mettere Mac=1 se sto lavorando sul file Mac
\newcommand{\ifMac}[2]{\ifnum\Mac=1 #1 \else #2 \fi}
\def\putps(#1,#2)(#3,#4)#5#6{\put(#3,#4){\epsffile{#6}}}

\def\Red  {}

\def\Black{}
\def\Blue {}
\newcommand{\hu}{{h_{\rm u}}}

\newcommand{\MS}{\overline{\hbox{\sc ms}}}
\newcommand{\fig}[1]{~\ref{fig:#1}}

\newcommand{\GeV}{\,{\rm GeV}}
\newcommand{\TeV}{\,{\rm TeV}}

\newcommand{\MGUT}{M_{\rm GUT}}

\newcommand{\One}{\hbox{1\kern-.24em I}}%%{\hbox{\bf 1}}%%
\newcommand{\NP}{Nucl. Phys.}

\newcommand{\PL}{Phys. Lett.}
\newcommand{\PR}{Phys. Rev.}
\newcommand{\mhu}{m_{h^{\rm u}}}

\newcommand{\eq}[1]{~(\ref{eq:#1})}

\def\circa#1{\,\raise.3ex\hbox{$#1$\kern-.75em\lower1ex\hbox{$\sim$}}\,}
\makeatletter
\def\art{\@ifnextchar[{\eart}{\oart}}
\def\eart[#1]#2#3#4#5#6{{\rm #2}, {\em #3 \bf #4} {\rm (#6) #5} ({\em #1})}
\def\hepart[#1]#2{{\rm #2, \em#1}}
\newcommand{\oart}[5]{{\rm #1}, {\em #2 \bf #3} {\rm (#5) #4}}
\newcommand{\y}{{\rm and} }
\newcounter{alphaequation}[equation]
\def\thealphaequation{\theequation\hbox to
0.6em{\hfil\alph{alphaequation}\hfil}}
\def\eqnsystem#1{
\def\@eqnnum{{\rm (\thealphaequation)}}
\def\@@eqncr{\let\@tempa\relax \ifcase\@eqcnt \def\@tempa{& & &} \or
  \def\@tempa{& &}\or \def\@tempa{&}\fi\@tempa
  \if@eqnsw\@eqnnum\refstepcounter{alphaequation}\fi
\global\@eqnswtrue\global\@eqcnt=0\cr}
\refstepcounter{equation} \let\@currentlabel\theequation \def\@tempb{#1}
\ifx\@tempb\empty\else\label{#1}\fi
\refstepcounter{alphaequation}
\let\@currentlabel\thealphaequation
\global\@eqnswtrue\global\@eqcnt=0 \tabskip\@centering\let\\=\@eqncr
$$\halign to \displaywidth\bgroup \@eqnsel\hskip\@centering
$\displaystyle\tabskip\z@{##}$&\global\@eqcnt\@ne
\hskip2\arraycolsep\hfil${##}$\hfil& \global\@eqcnt\tw@\hskip2\arraycolsep
$\displaystyle\tabskip\z@{##}$\hfil
\tabskip\@centering&\llap{##}\tabskip\z@\cr}
\def\endeqnsystem{\@@eqncr\egroup$$\global\@ignoretrue} \makeatother

\newcount\n

\begin{document}\twocolumn[
\centerline{hep-ph/9912301 \hfill IFUP--TH/63--99 \hfill OUTP 99--64P}
\vspace{5mm}

\Black
\vspace{0.5cm}
\centerline{\LARGE\bf\Red Are heavy scalars natural in minimal supergravity?}
\medskip\bigskip\Black
\centerline{\large\bf Andrea Romanino}\vspace{0.2cm}
\centerline{\em Department of Physics, Theoretical Physics}
\centerline{\em University of Oxford, Oxford OX1 3NP, UK}
\vspace{3mm}
\centerline{\large\bf Alessandro Strumia}\vspace{0.2cm}
\centerline{\em Dipartimento di Fisica, Universit\`a di Pisa {\rm and}}
\centerline{\em INFN, sezione di Pisa,  I-56126 Pisa, Italia}\vspace{1cm}
\Blue
\centerline{\large\bf Abstract}
\begin{quote}\large\indent
It has been recently claimed that very large values of a universal soft mass term $m_0$ 
for sfermions and higgs bosons
become natural when $M_t$ is close to $175\GeV$ if $\tan\beta\approx 10$.
We show that very large values of $m_0$ require accidental cancellations 
not guaranteed by experimental data or theoretical assumptions, and
consequently an unnatural fine-tuning of the parameters.

\end{quote}\Black
\vspace{0.5cm}]

\small

\noindent
While supersymmetric particles continue to be elusive,
it has been suggested that %, according to certain definitions of FT,
a very heavy universal scalar mass parameter
$m_0$ should be considered `natural',
so that all sfermions and non-SM higgses
could have multi-TeV masses above the LHC discovery reach.
The claim is
based on the observation that for values of the pole top mass $M_t$
around its experimental value~\cite{Mtexp}
%$M_t=(175\pm5)\GeV$ 
and for moderately large $\tan\beta$ the MSSM RGE equations for the
soft terms with minimal supergravity (mSUGRA) boundary conditions can exhibit a peculiar
behavior named `focus point' in~\cite{focus}: $\mhu(Q)$
(the soft mass term of the higgs $\hu$ coupled to up-quarks)
renormalized at a scale $Q\sim \TeV$
has a negligible dependence on its initial value at $Q\sim\MGUT$.

\begin{figure*}[t]
\begin{center}
\includegraphics{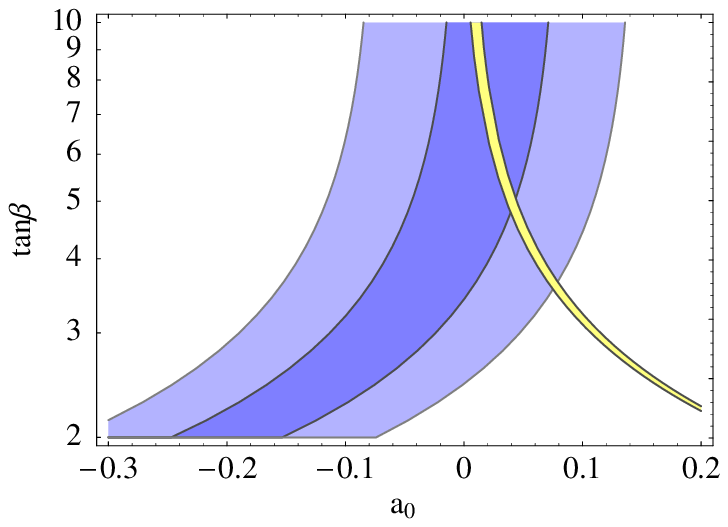}\hspace{5mm}
\includegraphics{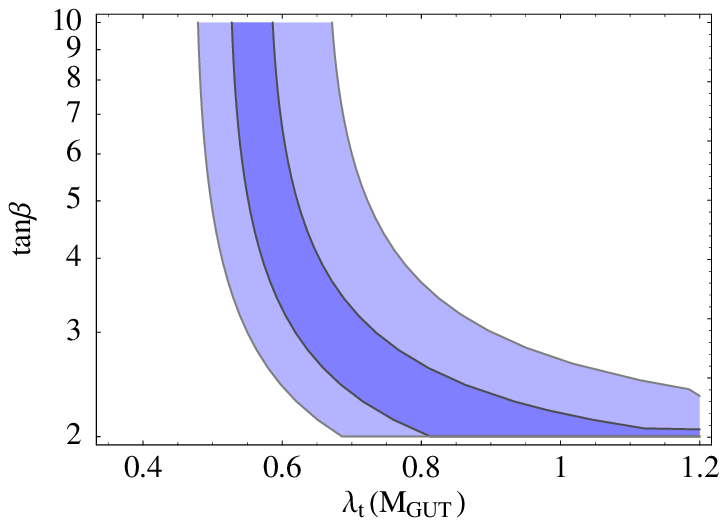}\hspace{5mm}~
%[width=7cm,height=5cm]
% %\begin{picture}(16,6)
% %\putps(0,0)(0,0){a0tb}{a0tb.eps}
% %\putps(8,0)(8,0){fban}{ban.eps}
% %\end{picture}
\caption[SP]{\em The lighter regions show the uncertainties on the
$(a_0,\tan\beta)$ (fig.\ 1a) and $(\lambda_t(\MGUT),\tan\beta)$ (fig.\
1b) plane induced by $1\sigma$ uncertainties on $\alpha_3$ and $M_t$
and by the uncertainty on the sparticle spectrum.  In fig.~1a we have
fixed $m_0=3\TeV$.  Despite the top mass is precisely known, the
values of $a_0$ and $\lambda_t(\MGUT)$ are still quite uncertain.
Even if $M_t$ were perfectly measured, the uncertainties would be only
partially reduced (inner darker regions).  Only if $a_0$ lies inside
the thin light region the correction to $M_Z^2$ proportional to
$m_0^2$ is smaller than $10 M_Z^2$.\label{fig:fltG}}
\end{center}\end{figure*}

It is easy to understand what a `focus point' is:
RGE effects trigger 
electro-weak symmetry breaking (EWSB)
by converting a positive value of $\mhu^2(\MGUT)$ into a negative
value of $\mhu^2(Q)$ if the top Yukawa coupling is sufficiently large.  On the contrary, $\mhu^2(Q)$ remains positive if
the top Yukawa coupling is too small.  Consequently, $\mhu^2(Q)$
vanishes for some appropriate intermediate value
$$\lambda_t\sim4\pi/\ln^{1/2}(\MGUT^2/M_Z^2)\sim 1$$ of
the top Yukawa coupling.
Starting with mSUGRA boundary conditions (universal gaugino masses $m_{1/2}$ and universal
scalar masses $m_0$ at $\MGUT\approx 2~10^{16}\GeV$; for ease of illustration we assume a vanishing $A_0$-term),
we can write
$$\mhu^2(Q)= a_0\cdot m_0^2+a_{1/2} \cdot m_{1/2}^2.$$ In absence of
radiative corrections $a_0=1$ and $a_{1/2}=0$.  For an appropriate
value of $\lambda_t(\MGUT)$ close to $1/2$ the coefficient $a_0$
vanishes and a large $m_0$ can coexist with a small $M_Z$.  Such a
cancellation had already been noticed when the notion of fine-tuning
(FT) had been introduced~(see fig.~1a of \cite{FT}).  If the scalar
soft terms are non-universal the value of $\lambda_t$ giving the
analogous cancellation is different.  An experimentally acceptable top
mass $M_t\approx v\lambda_t \sin\beta$ can be obtained with an
appropriate choice of $\tan\beta$.  With universal soft terms, $a_0$
can vanish for moderately large values of $\tan\beta$~\cite{focus} (a
regime where $\sin\beta\approx 1$ is fixed).

Unfortunately such cancellation, even if taking place, would not allow to
improve the present unsatisfactory `naturalness status' of mSUGRA
models~\cite{CEP,nat}, mainly determined by the radiative
contribution to $M_Z^2$ proportional to the squared gluino mass
$M_3^2$, a few times larger than $M_Z^2$ itself. On the contrary, the
$m_0^2$ contribution to $M_Z^2$ is not problematic so that suppressing
it would not help~(see appendix~\ref{nat}). However, the possibility
that mSUGRA does not become less natural when $m_0\gg M_Z$
would certainly have implications for
model-building and experiments.

Is a cancellation between the $m_0^2$ contribution to $M_Z^2$ and the
radiative corrections to it more `natural' than a cancellation between
different soft terms?  A FT analysis says that a cancellation in the
$m_0^2$ contribution allows to have a heavy $m_0$ without a large FT
of the soft terms, but with a large FT of the couplings (mainly
$\lambda_t$ and $\alpha_3$).  The FT associated with the couplings is
sometimes included, omitted or neglected in the various definitions of
FT employed in the literature.  This choice is usually irrelevant
(because the fine-tuning with respect to the $\mu$-term is often the
strongest one), but not in this case.
The FT-parameter used in~\cite{focus} does not include the FT
associated with the couplings.
In the following, we will discuss why
and how it has to be included,
making too large values of $m_0$ unnatural.

The real issue does not consist of computing a number that should
quantify ``how much we like'' the cancellation necessary to have a
large $m_0$.  The problem of `unnatural situations' (like a strong
accidental cancellation) is that they are unlikely, because they
happen only in a small percentage of the available parameter space.

Consequently, in order to assess if $m_0\gg M_Z$ is natural in minimal
supergravity with $\tan\beta\approx 10$, what we should actually
determine is
whether the experimental determination of $M_t$ implies that the
necessary cancellation is happening  i.e.\
if the coefficient $a_0$
%$$M_Z^2 \approx -2|\mu|^2-2 \mhu^2(M_Z) =  -2(a_0  m_0^2+a_{1/2}
%m_{1/2}^2 + |\mu^2|) $$ 
is forced to be much smaller than its typical value at $\tan\beta\approx 10$
\begin{equation}\label{eq:a0typ}
|a_0(M_t,\alpha_i(M_Z),\hbox{sparticle spectrum})|\sim 0.2.
\end{equation}
The answer is no.
The experimental uncertainty on $M_t$, on $\alpha_3(M_Z)$ and on the
sparticle spectrum induces an uncertainty on $a_0$ comparable to
its `typical' value, $|a_0|\sim 0.2$, due to the strong sensitivity of
$a_0$ to these parameters.

One way of understanding such a strong sensitivity is to neglect the
threshold corrections to $a_0$ and to express the dependence of $a_0$
on the EW parameters through an integral involving the top Yukawa
coupling renormalized at energies $\mu$ higher than the EW scale:
$a_0\approx 1-\frac{3}{2} \rho$, where
$$\rho\equiv1-\exp\bigg[-6\int_{\ln m_0}^{\ln M_{\rm GUT}}
\lambda_t^2(\mu)\frac{d\ln \mu}{8\pi^2}\bigg].$$ While $M_t$ and
$\alpha_3(M_Z)$ are known with few $\%$ uncertainty, there is a larger
uncertainty on $\lambda_t(\mu)$: unknown sparticle threshold
corrections affect the value of $\lambda_t$ just above the SUSY
breaking scale; the running up to higher scales depends on the gauge
couplings (also affected by unknown sparticle threshold corrections)
amplifying the uncertainties in $\lambda_t$.  Of course, by solving
the RGE equation for $\lambda_t$~\cite{RGEMSSM}, $\rho$ can be written
in terms of the value of $\lambda_t$ renormalized at any scale between
$m_0$ and $\MGUT$: for example
$$\rho = \frac{\lambda_t^2(m_0)}{E/6F}
=\frac{1}{1+[6F\lambda_t^2(\MGUT)]^{-1}}$$ where $E$ and $F$ are
functions of the gauge couplings $g_i$ defined as~\cite{RGEMSSM}
$f_i(Q)\equiv \alpha_i(\MGUT)/\alpha_i(\mu)$, $E(\mu)\equiv
f_1^{13/99}f_2^3f_3^{-16/9}$, $E\equiv E(m_0) \approx 11.$ and
\begin{equation}\label{eq:F}
F\equiv \int_{\ln m_0}^{\ln M_{\rm GUT}} E(\mu)\frac{d\ln
\mu}{8\pi^2}\approx 1.5.
\end{equation}
Writing $a_0$ in terms of $\lambda_t(m_0)$ we can estimate the
uncertainty on $a_0$\footnote{Using $\lambda_t(\MGUT)$ we would
get the same uncertainty on $a_0$, comparable to $a_0$.
Therefore we do not agree with
J.L Feng, K.T. Mathcev and T. Moroi, hep-ph/0003138.} due to the uncertainty on the couplings as
$$\delta a_0 \approx-2.5\, \delta \lambda_t(m_0) + 2.2\, \delta
g_3(m_0) + \cdots.$$ Even if $\lambda_t$ and $g_i$ were measured with
negligible error at the $Z$-scale, unknown threshold corrections would
still induce a $\sim 0.1$ uncertainty on $a_0$.
%(In absence of any threshold correction the unification prediction for
%$g_3$ would be few standard deviations too high).
We illustrate this uncertainty in fig.\fig{fltG}a,
where we show the allowed region of the
($a_0,\tan\beta$) plane
corresponding to
$$M_t=(175\pm  5)\GeV,\quad \alpha_{3}(M_Z)_{\MS}=0.120\pm0.003$$
$$m_0=3\TeV,\qquad 300\GeV<M_3<1\TeV$$ (lighter region).  The inner
darker region has been plotted assuming $M_t=175\GeV$ in order to show
that a significant uncertainty on $a_0$ would be present even if $M_t$
were perfectly known.  Shown are also the values of $a_0$ for which
the $m_0^2$ contribution to $M^2_Z$ is smaller than $10 M^2_Z$. For
completeness, in fig.\fig{fltG}b we show the allowed regions of the
($\lambda_t(\MGUT),\tan\beta$) plane corresponding to the same
parameter ranges, except for $m_0$, now varied between $200\GeV$ and
$1\TeV$\footnote{We have varied the range because knowing the allowed
range the top quark Yukawa coupling renormalized at the unification
scale as function of $\tan\beta$ is also interesting for
lepton-flavour violating signals of supersymmetric
unification~\cite{BH}.
Fig.~1b shows in a less direct but more precise way than fig.~1a
that there is no evidence for a very small value of $a_0$.}.

\smallskip

To summarize, fig.\fig{fltG} shows that there is no experimental
evidence that $\lambda_t$ is very close to the value that gives
$a_0=0$ --- i.e.\ that a cancellation is suppressing the $m_0^2$
contribution to the $Z$ mass.  Although such a suppression is not
excluded, it would require a FT of the relevant parameters inside
their experimental ranges. As a consequence, $m_0$ can be heavy only
if some cancellation is forced: between $m_0^2$ and the radiative
corrections to it (by fine-tuning the couplings), or between $m_0^2$
and other soft terms (for example by fine-tuning the $\mu$ term), or
both.  In both cases a significant cancellation is unlikely and
therefore unnatural\footnote{If this conclusion were not true, any
supersymmetric model with very heavy sparticles could be made
`natural' provided that the soft terms depend on unmeasured couplings.
Even the quantum corrections to the higgs mass in the
non-supersymmetric SM could be made `naturally' vanishing by choosing
an experimentally allowed appropriate value of the SM quartic higgs
coupling.}.

\medskip

Having explained our main point, we rediscuss it in a more quantitative way.
In order to compute the
naturalness upper bound on $m_0$ we have to estimate how unlikely is
the cancellation necessary to allow large values of $m_0$.

The FT parameters quantify how sensitive is $M_Z$ with respect to
variations of the parameters.  Sensitivity and naturalness are however
two different things~\cite{FTcritica,GMFT,nat}.  Nevertheless, $1/{\rm
FT}$, if much smaller than one and if divided by the `total allowed
parameter space', gives a rough measure of the percentage of the
allowed parameter space where a certain cancellation
happens~\cite{GMFT}.  In absence of a theoretical justification, very
strong cancellations happen only in very small corners of the
parameter space and are consequently very unlikely.  To estimate how
unlikely are the cancellations that allow a large $m_0$, we must
therefore include the FT with respect to each relevant parameter $\wp$
and normalize it with respect to their experimentally allowed range
$\Delta\wp$~\cite{GMFT}.  More precisely, we will compute
$$\Delta(\wp) \simeq
%\left|\frac{\Delta \wp}{M_Z^2}\frac{\Delta M_Z^2}{\Delta  \wp}\right|
\left|\frac{\Delta \wp}{M_Z^2}\frac{\partial M_Z^2}{\partial
\wp}\right| \quad\hbox{instead of}\quad {\rm FT}(\wp)
\simeq\left|\frac{\wp}{M_Z^2}\frac{\partial M_Z^2}{\partial
\wp}\right|
$$
for each parameter $\wp$. We will then combine different FTs in the
`usual' way:
$$ \Delta = \mbox{max}_\wp [ \Delta (\wp) ]. $$
although, since we want to estimate the probability that two different and almost
independent cancellations could occur,
it would be safe to multiply the FT parameters relative to the two cancellations,
obtaining stronger bounds.

While $m_0$ was the
only parameter considered in~\cite{focus}, we consider in addition the
FTs with respect to variations of $M_t$, $\alpha_3(M_Z), \ldots$ in their experimental ranges.
%and the parameters determining the thresholds for fixed $m_0$.
At present, $\Delta (M_t)$ is
actually the only relevant FT besides $\Delta(m_0)$. We assume that the uncertainty on $m_0^2$ is
comparable to $m_0^2$ (so that $\Delta(m_0^2)\approx {\rm
FT}(m_0^2)$), and we (optimistically) assume a $\Delta\wp=0.1$ total uncertainty range
on $\wp=\lambda_t(\MGUT)$\footnote{This is the minimal uncertainty
that would be obtained if $M_t$ were known with negligible error;
including the present error on $M_t$ would make $\lambda_t(\MGUT)$
more uncertain, see fig.\fig{fltG}, strengthening our
conclusions.}. As observed above, one can exploit the relation between
$M_t$ and $\lambda_t(\MGUT)$ and use $\Delta (\lambda_t(\MGUT))$
instead of $\Delta (M_t)$. The two possibilities are equivalent in the
limit in which the uncertainty on $M_t$ is the dominant one.
If the error on $M_t$ will be reduced down to a negligible level,
$\Delta(\lambda_t(\MGUT))$ will still take into account
(some of) the FT associated to, e.g., $\alpha_3$
so that our conclusions will still hold.

\begin{figure}[t]
\begin{center}
\includegraphics{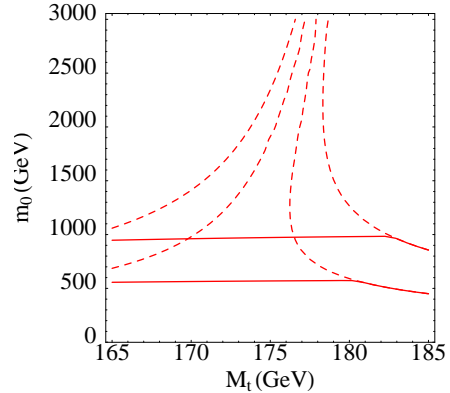}
% \begin{picture}(8,7)
% \putps(0,0)(0,0){fFT}{FT.eps}
% \end{picture}
\caption[SP]{\em Naturalness upper bounds on $m_0$ as a function of
$M_t$ ($\Delta<10, 30$) before (dashed lines) and after (continuous lines) 
having properly taken into account the uncertainty on the relevant parameters.
\label{fig:FT}}
\end{center}\end{figure}

Let us discuss analytically the magnitude of $\Delta
(\lambda_t(\MGUT))$.  If the experimental measure of $M_t$ implied
that $|a_0|\ll 1$, our FT-like parameter would consider as natural
very large values of $m_0$.  The variation of $a_0$ with
$\lambda_t(\MGUT)$ is however sufficiently strong to disfavour such a
possibility:
$${\rm FT}(\lambda_t(\MGUT))=\frac{36 F\cdot\lambda_t^2(\MGUT)}{[1+16
F\cdot\lambda_t^2(\MGUT)]^2}\frac{m_0^2}{M_Z^2}$$ where $F$ has been
defined in\eq{F}.  For $\lambda_t(\MGUT)$ close to the value where the
cancellation in $a_0$ takes place,
$$\Delta(\lambda_t(\MGUT))\approx \frac{0.1}{\lambda_t(\MGUT)}\,{\rm
FT}(\lambda_t(\MGUT))\approx 0.25\frac{m_0^2}{M_Z^2}.$$ 
The effect of taking $\Delta(\lambda_t(\MGUT))$ into account is shown
in fig.\fig{FT}.
%We have taken into account the variation of the $\lambda_t(\MGUT)$
%range with $m_0$, 
We have assumed fixed values for the gaugino masses and for the gauge
couplings.  For heavy $m_0$ there is a small portion of parameter
space (limited by the dashed lines) where $\Delta(m^2_0)<10,30$. As
explained, the smallness of this region means that there is a
significant FT with respect to some other parameter.  In fact this
regions disappears when $\Delta(\lambda_t(\MGUT))$ (solid line) is
taken into account.

\medskip

In conclusion, very heavy values of $m_0$ require an unnatural FT of
the relevant parameters inside their present experimental range.
%Using $\lambda_t(\MGUT)$ as a parameter in place of $M_t$ makes the computation simpler
%but of course does not affect the result.
We have used the FT-like parameter introduced in~\cite{GMFT} and
repeated the computation in appendix~\ref{m0:nat} using the more
accurate technique presented in~\cite{nat}.  With respect to this
problem both criteria are {\em less} restrictive than a `na\"\i{}ve'
complete FT analysis.  In both cases the result is that too large
values of $m_0$ are unnatural, as it can simply be seen by inserting a
{\em typical\/} value of $|a_0|$ and the preferred confidence level on
unlikely cancellations (for example ${\rm FT}_{\rm
lim}\circa{<}1/10\%$) in the naive bound $m_0^2\circa{<}{\rm FT}_{\rm
lim}M_Z^2/|2a_0|$.  Since $a_0$ is typically small, eq.\eq{a0typ}, one
obtains the usual weak naturalness bound on $m_0$, well above all
present accelerator bounds, but not above $1\TeV$.  Due to the
smallness of $|a_0|$, the naturalness upper bound on $m_0$ has almost
no impact on the `naturalness status' of mSUGRA models, as discussed
in appendix~\ref{nat}.

\paragraph{Acknowledgements}
The work of A.R. was supported by the TMR Network under the EEC
Contract  No. ERB\-FM\-RX--CT960090.
We thank R. Barbieri, G. Giudice and G.G. Ross for useful discussions.

\appendix

\section{Naturalness bound on $m_0$}\label{m0:nat}
As said naturalness disfavours heavy $m_0$ because very strong
cancellations (either between different soft terms, or between the
tree level $m_0^2$ term and the radiative corrections to it) are
needed in order to accommodate very large values of $m_0$.  Setting a
naturalness upper bound on $m_0$ amounts to estimate how unlikely is
the required cancellation in the light of our experimental and
theoretical knowledge.

 If we assign to the parameter space an arbitrary probability
 distribution function (pdf) we can compute the probability of any
 event,
for example of the required cancellation.
The pdf is however totally arbitrary in absence of
 experimental data.  This same assumption (the choice of an arbitrary pdf, called
 `Bayes prior' in statistical inference) is the crucial ingredient that
 allows to convert experimental data into measured ranges of
 fundamental parameters, like the top mass.  Starting from an arbitrary
 pdf and using simple properties of probability, it is possible to
 follow how experimental data modify the probability of different
 values.  When experimental information is sufficiently strong, the
 final pdf does not depend on the arbitrary pdf needed to start with.
 This is why we can today assume that the pole top mass is distributed
 according to a $175\pm 5$ gaussian. 

 Since the soft terms are totally unknown we assume some broad pdf for them.
Our results have only a mild dependence on the pdf,
unless some crazy pdf is chosen.
 Since $M_Z$ (that is one combination of soft terms) has been already
 measured with a practically infinite precision, it is simpler to take
 this experimental constraint into account with the procedure used in~\cite{nat}:
 we assume a 
probability distribution for the dimensionless ratios of the 
soft terms, and compute the overall scale of soft terms from the
EWSB condition.  Since in this way we never specify how heavy are the
sparticles, the connection of this procedure with naturalness is quite
transparent.

Sampling all parameters, like $M_t$ and $m_0/m_{1/2}$,
according to their assumed pdf, we estimated~\cite{nat} that only in
$p\sim5\%$ of the cases a cancellation in the EWSB conditions
generates sparticle masses above all experimental bounds in mSUGRA.
In order to set upper bounds on $m_0$ we repeat the analysis
in~\cite{nat}, but without averaging $p$ over the distribution of
$m_0/m_{1/2}$: we here compute $p$ as function of
$m_0/m_{1/2}$\footnote{ We could also study $p$ as function of
$m_0/M_Z$.  However $m_0\gg M_Z$ is possible either because $|a_0|\ll
1$, or due to a cancellation between different soft terms.  We study
$p(m_0/m_{1/2})$ rather than $p(m_0/M_Z)$ because we here want to
concentrate our attention on the first possibility.  Bounds on
$m_0/M_Z$ have a more direct impact on phenomenology.  Bounds on
$m_0/m_{1/2}$ have a more direct impact on theoretical attempts of
predicting $m_0/m_{1/2}$.}
at fixed $\tan\beta=10$.  We find that $p(m_0/m_{1/2})$ has a maximum
at $m_0\sim 3 m_{1/2}$, decreases when $m_0\ll m_{1/2}$ (because too
small values of $m_0$ give light right-handed sleptons) and becomes
negligibly small when $m_0\gg m_{1/2}$ (more precisely when
$m_0\circa{>}3 M_3$).  We again conclude that values of $m_0$
significantly above $1\TeV$ require very unlikely cancellations in the
EWSB condition.  A certain minimal amount of cancellation is however
required even for $m_0$ below $1\TeV$ in order to accommodate
experimental bounds, as recalled in appendix~C.

\section{Heavy $m_0$ and the naturalness problem}\label{nat}
The $Z$ mass is given, as function of the soft terms,
by a potential minimization condition that in mSUGRA with vanishing $A_0$
and large $\tan\beta\approx 10$ can be approximated as
\begin{equation}
M^2_Z =-2(a_0 m^2_0 + a_{1/2} m_{1/2}^2+ \mu^2).\label{eq:MZ}
\end{equation}
One important success of supersymmetry is the prediction that RGE
effects typically induce negative $a_i$ coefficients, thus
establishing a direct link between SUSY-breaking and EW-breaking.
This nice feature is however due to $\lambda_t$ and $g_3$
interactions: SUSY breaking most naturally induce a non vanishing
$Z$-boson mass comparable to the gluino and top-squark masses, that
are typically heavier than the other non coloured sparticles.  On the
contrary experiments now tell that the $Z$ boson is lighter than
(almost) all sparticles.  This kind of naturalness problem manifests
itself in eq.\eq{MZ} if the bounds on sparticle masses imply that the
single contributions to $M_Z^2$ are much larger than $M_Z^2$ itself.
What happens is that the $m_0^2$ contribution gives no
problems, while the $m_{1/2}^2$ term gives an unpleasantly large
contribution~\cite{CEP,nat} to $M_Z^2$, that can be canceled by the $\mu^2$
term.

The $m_0^2$ contribution does not pose naturalness problems because
the experimental bound on $m_0$ is weak ($m_0$ could even be zero),
and because the coefficient $a_0$ is typically small, $-a_0<1/3$.
The particular structure of the SUSY RGE protects the $m_0^2$ contribution from
QCD corrections, that instead affect the $m_{1/2}^2$ contribution.
This well known fact can be easily understood with the techniques of~\cite{GR}.

The $m_{1/2}^2$ term is problematic because
it has a large coefficient $a_{1/2}\sim -(3\div 5)/2$ and because
LEP and Tevatron experiments provide significant lower bounds on $m_{1/2}$.
The $m_{1/2}^2$ contribution to $M_Z^2$ is approximately given by
$$\frac{M_Z^2}{(91\GeV)^2}=(5\div 11)(\frac{M_3}{290\GeV})^2+\cdots$$
where $M_3\approx 2.5 m_{1/2}$ is renormalized at $Q=500\GeV$ and lower values in the given range
can be obtained for higher $\tan\beta$ and lower $\lambda_t(\MGUT)$.
The LEP limit on the chargino masses gives rise, due to our assumption of gaugino mass unification,
to a strong but indirect bound on the gluino mass, $M_3\circa{>} 290\GeV$.
Abandoning gaugino mass unification only the Tevatron direct bound on the gluino mass applies
($M_3\circa{>}(180\div 280)\GeV$, depending on the squark spectrum)
so that the situation can be partially improved~\cite{nonGUT,nat}.
%$$\frac{M_Z^2}{(91\GeV)^2}=(4.2\div 7.5)(\frac{M_3}{250\GeV})^2+\cdots$$
%$$\frac{M_Z^2}{(91\GeV)^2}=(2.7\div 4.8)(\frac{M_3}{200\GeV})^2+\cdots$$
%On the contrary, since the $m_0$ term does not lead to naturalness problem,
%the minumum amount of fine-tuning cannot be decreased if $b_0=0$
%and/or the universality 
%hypothesis on the sfermion masses is relaxed.
The value of $m_0$ has only a small indirect impact on the naturalness problem:
since $\tan\beta$ is determined by minimizing the potential,
a moderately large $m_0$ allows to naturally obtain
the moderately large values of $\tan\beta\sim 10$
for which the $m_{1/2}^2$ problem is minimized~\cite{CEP2}.

\end{document}